# Data-driven quantitative analysis of an integrated open digital ecosystems platform for user-centric energy retrofits: A case study in Northern Sweden

Bokai Liu[a],*, Santhan Reddy Penaka[a], Weizhuo Lu[a], Kailun Feng[a], Anders Rebbling[a], Thomas Olofsson[a]

[a] *Department of Applied Physics and Electronics, Umeå University, 901 87 Umeå, Sweden*

**Abstract**

This paper presents an open digital ecosystem based on a web-framework with a functional back-end server for user-centric energy retrofits. This data-driven web framework is proposed for building energy renovation benchmarking as part of an energy advisory service development for the Västerbotten region, Sweden. A 4-tier architecture is developed and programmed to achieve users' interactive design and visualization via a web browser. Six data-driven methods are integrated into this framework as backend server functions. Based on these functions, users can be supported by this decision-making system when they want to know if a renovation is needed or not. Meanwhile, influential factors (input values) from the database that affect energy usage in buildings are to be analyzed via quantitative analysis, i.e., sensitivity analysis. The contributions to this open ecosystem platform in energy renovation are: 1) A systematic framework that can be applied to energy efficiency with data-driven approaches, 2) A user-friendly web-based platform that is easy and flexible to use, and 3) integrated quantitative analysis into the framework to obtain the importance among all the relevant factors. This computational framework is designed for stakeholders who would like to get preliminary information in energy advisory. The improved energy advisor service enabled by the developed platform can significantly reduce the cost of decision-making, enabling decision-makers to participate in such professional knowledge-required decisions in a deliberate and efficient manner. This work is funded by the AURORAL project, which integrates an open and interoperable digital platform, demonstrated through regional large-scale pilots in different countries of Europe by interdisciplinary applications.

## 1. Introduction

Energy-efficient retrofitting measures (EEMs) to minimise energy use in individual buildings and building stock are necessary due to fast global energy demand growth and climate change awareness [1] [2] [3]. Energy efficiency in buildings is crucial now due to rising European energy prices. The buildings sector in Europe, the main focus has been on improving energy efficiency in new buildings [4]. However, since more than 30% of European buildings are older than 50 years and 75% of them are energy inefficient, EEMs must be prioritised [5]. John Dulac, analyst at International Energy Agency, predicts that adopting EEMs in new buildings extensively across Europe will still not be enough, given the millions of existing old buildings [6]. Sweden built 1.4 million dwellings between 1946 and 1975 as part of a national programme. Since many will be over 50 years by 2030, many need renovation. They also have less thermal insulation and heat recovery ventilation [7]. To satisfy energy efficiency standards, about 50,000 houses must be renovated annually. To speed up renovations, building owners need decision-making knowledge [8] [9] [10]. Though Swedish house owners want to renovate, there are no dedicated platforms in Sweden that can identify user-specific renovation needs and provide insights during the deliberation, planning, and finalisation phases of the decision-making process.

* Corresponding author: bokai.liu@umu.se

Decision support systems (DSS) or expert systems can cope with this situation, as they usually involve computer-assisted procedures to help users solve undetermined problems through data and optional models [11]. Among them, the energy renovation system specifically analyzes the energy performance of various residential buildings through input, and gives users energy consumption evaluation through the screening and analysis of existing database. On the one hand, the popularity of DSS or expert systems lies in homeowners' needs for energy retrofits in their houses, which makes energy advisory a good guide [12]. In addition, with the development of data-driven techniques, a large operational capacity can be used for analysis and modeling based on database [13]. Developments and improvements in computer science have enabled scientists or researchers to access vast amounts of information and create cutting-edge analytical tools. On the other hand, these novel tools should be user-centric and flexible.

User-centric, or as it is also called, user-centered friendliness, means that there is a modular framework for different users, which is simple and effective with a strong generalization ability [14]. Flexibility suggests that the system should allow users to operate it stably and consistently across various devices without being constrained by the system environment or hardware configuration. Based on the above objective needs, the web-based platform has become a broad and effective solution. There are many studies on web-based engineering application methods. Suprun et al. provided a web-based application that can achieve randomization for microplate experiments [15]. Stephen Greaves et al. propose A Web-Based Diary and Companion Smartphone app for Travel/Activity Surveys, which improves the accuracy of trip reporting [16]. Alessandro Sopegno et al. create A web mobile application for agricultural machinery cost analysis, it can support the decisions on whether to purchase/rent/hire a new equipment, select the economical appropriate cultivation system [17]. Efrén Fitz-Rodríguez et al. design A web-based application for dynamic modeling and simulation of greenhouse environments under several scenarios [18]. In our previous research, we develop a web-based app for computationally expensive models in materials design [19] [20]. In addition to user-friendliness and interaction design, a "good" platform requires a versatile server and an effective data modeling tool, which requires data-driven techniques [21] [22] [23] [24]. For example, through data mining of existing databases, a large amount of data screening and analysis can be used to generate a model that can discover mechanism. We call this data-driven technology the fourth paradigm [25]. For the construction of database and ecosystems, related algorithms and functions can be deployed and implemented, which plays a vital role in realizing user-centered design.

Based on the above existing technological development and social needs, the European Union launched a project: *Architecture for Unified Regional and Open digital ecosystems for Smart Communities and Rural Areas Large scale application (AURORAL)*. AURORAL is an EU Horizon 2020 project with 25 organizations in the consortium from 10 EU nations (*https://www.auroral.eu/*). The objective of the project is to establish an interoperable, open, and digital middleware platform for rural ecosystem services in Europe, hence enabling equal opportunity for rural and urban Europeans. Cross-domain applications are used to demonstrate the AURORAL middleware platform across eight large-scale pilot regions in Europe, including the Västerbotten region of Sweden. Västerbotten is the second largest region in Sweden and is comprised primarily of sparsely populated rural areas [26]. Due to the aging of existing buildings, there is a demand for decision-making knowledge for energy-efficient building renovation in the region. Consequently, the purpose of this pilot is to offer energy advising services to building owners in all 15 municipalities. The developed benchmark model can reach beneficiaries in remote places via a digital platform such as the AURORAL middleware because internet connectivity in the region is sufficient.

In this context, the authors present an open digital ecosystem, a web-based framework platform built upon a functional back-end server that utilizes data-driven methods. These functions are achieved through uncertainty modeling and quantitative analysis of database data. Our contributions to the expert system in energy retrofits are as follows: (1) We introduce a systematic framework that can be applied in energy advisory with data-driven approaches; (2) The web-based platform is user-centric and flexible for operation; (3) The feedback and visualization are integrated into the framework with preset algorithms. This computational framework is designed for stakeholders who would like to do the preliminary consulting before the energy renovation.

This article is organized as follows. In the next section 2, we describe the overall ecosystem web-based platform architecture with 4-tiers model. We introduce our quantitative analysis methodology in Section 3, which is based on variance method sensitivity indices approaches and data-driven modeling. Subsequently, we discuss the results from our implementation on Swedish pilot before we conclude our manuscript in Section 4.

## 2. The architecture for ecosystem platform

The focus of the digital ecosystem is bringing additional value to users by optimizing the data and workflows of different internal sectors, tools, systems as well as users, suppliers and external partners. Normally, the platform should eliminate the barriers in the usage and enable each participant in the ecosystem to access the most advanced functions and systems to meet their individual needs, which we call it 'user-centricity' [27].

These ecosystems provide users with a unified, easy-to-use system that delivers value through a variety of services, products, and insights. In this process, the central position is the platform, whose role is providing a place for other stakeholders to interact and help them complete the connection and matching. This interaction among stakeholders is carried out through the platform, including users, operators, and suppliers.

Due to the various above-mentioned requirements, web-based applications can be considered as the most popular digital ecosystem designs. First of all, this approach can effectively achieve user-friendliness. Users do not need to pay attention to the installation, debugging, and update/maintenance of programs, nor do they need specific system platforms and equipment. They can focus more on solving customized engineering needs. As for the developers, all development and maintenance in project are concentrated on one platform, which can reduce the uncertainty of specific engineering problems [19].

The web server is deployed on the Swedish Science Cloud (SSC), and users can access it through a permanent link: *http://130.238.28.85:8080/question.html*.

*2.1. System Architecture*

For the above reasons, we adopt a web-based application for user interaction design, which applies a multi-layer service-oriented user-server architecture model (4-tier architecture). It includes a front end for users and a back end server according to a interaction design theory. They are the user layer and web layer integrated in the front end interface, and the application layer and dataset layer embedded in the back end server [28].

The ecosystem can be accessed through a web browser and it is also compatible with different browsers. The main webpage is shown in Fig 1. Users can freely use the functions by logging into the exact domain address. While users submits tasks, the datastream goes through the backend server where the tasks are calculated by the preset algorithm. The result later is displayed visually after the computation through the front-end browser. Users can view and check their results via interactive display in real time. When errors or problems occur, corresponding warnings will show up. The whole 4-tier user-server architecture can be found in Fig 2

*2.1.1. User tier*

At the user tier, our main consideration is to create a user-centered interactive platform with a UI page that allows users to operate in a friendly manner. Users refer to decision-making based on the results of the interaction and the information they obtain. For general decision-making problems, we adopt Theory of Planned Behavior (TPB) as our applied model, explaining and designing user-tier steps [29].

The Theory of Rational Action (TRA) as a foundational theory was expanded by Ajzen in 1991 to form the Theory of Planned Behavior (TPB) to explain behavior [30]. This is achieved by adding the concept of perceived ease or difficulty of performing a behavior, hence TPB is also known as perceived behavioral control. The TPB proposes that behavioral intentions are influenced by attitudes toward behavior, subjective norms, and perceived behavioral control as proposed by TRA. According to this theory, 3 key components like subjective norms, attitudes, and perceived behavioral control are manifestations of beliefs currently in the mind [31]. The theoretical structure is shown in Figure 3.

TPB-based models often use hypotheses to confirm the effectiveness in the study. Corresponding underlying hypotheses are also proposed in our study to provide additional explanations for the design.

- **Hypothesis 1 (Subjective Norms)**:

  Subjective norms (N) represent the degree to which people perceive the importance and influence of the target behavior on employee value expectations. According to the TPB, these norms will have an effect on the individual's intention (I) to perform the intended target behavior (energy retrofits).

  Through the comparison of research related to energy retrofits, we realize an important subjective norms for energy retrofitting by stakeholders is that they find their own Energy Use Intensity (EUI) is higher than that of their neighbors. This comparison has led stakeholders to start paying attention to the energy efficiency of their homes [7]. From this situation, we speculate that if there is a benchmark to stakeholders as a reference, they will have a greater motivation to make decisions. Based on these results, we infer that energy retrofit behavior may be influenced by similar social influences and hypothesize that:

  **H1**: Individually important people's perceptions of energy retrofits (subjective norms) would be positively correlated with benchmark reference.

- **Hypothesis 2 (Attitude)**:

Figure 1: The main page of this web-based application

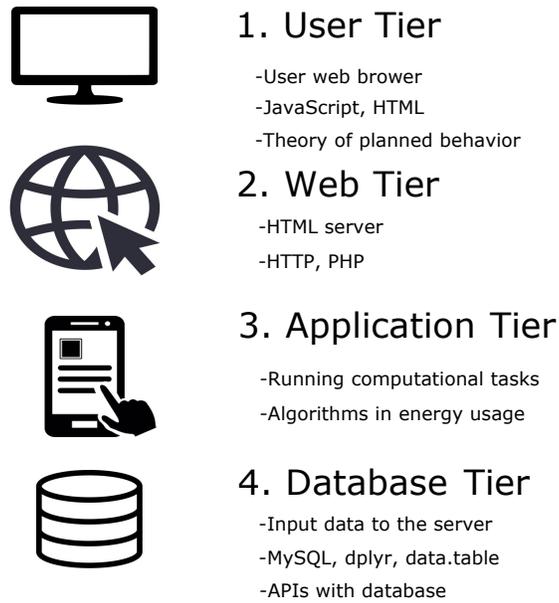

Figure 2: The Architecture of the web-based system

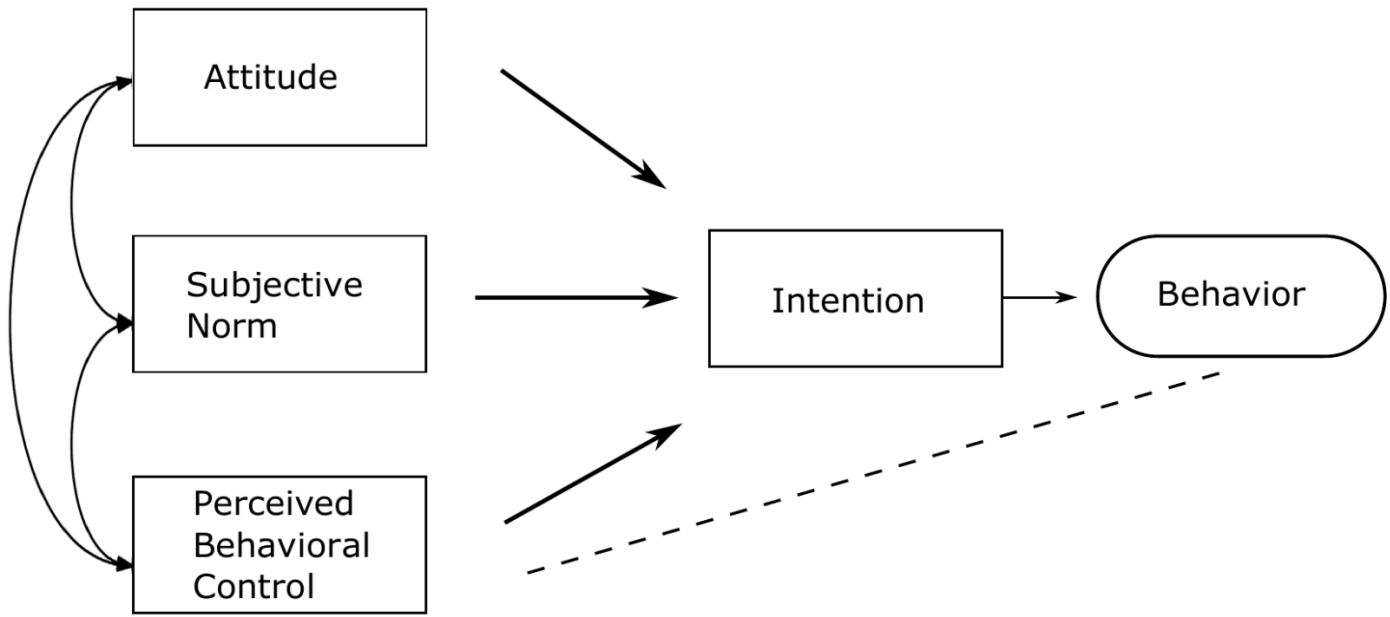

Figure 3: Theory of planned behavior

According to TPB, attitude (A) is a psychological evaluation of behavior, such as good/bad, positive/negative, etc. The TPB asserts that attitudes will directly affect an individual's intention to perform behaviors such as discomfort-retrofitting. A study by Shoaib Azizi et al. provides direction for the impact of subjective norms on energy retrofits. Their work suggests that the benefits stakeholders are interested in may have less to do with energy and more to do with indoor comfort, but such decisions ultimately lead to energy savings [32]. We infer that energy retrofit behavior may be similarly influenced by attitudes, assuming that:

**H2**: Perception of the value/effectiveness of energy retrofitting (attitude) would be positively correlated with intent to indoor thermal comfort.

- **Hypothesis 3 (Perceived Behavioral Control)**:

According to the TPB, perceived behavioral control (C) represents an individual's perception of the degree of control they have over performing a desired behavior. The TPB proposes that perceived behavioral control is not only related to intentions, but also to the actual performance of desired behaviors (eg, feasibility of retrofitting).

Finding resulting from research, there is a strong relation between the availability of retrofit facilities (an expression of perceived behavioral control) and energy retrofit intentions and energy retrofit activities. The implementation of energy retrofits usually requires a lot of information about services, products and processes [32]. From an information standpoint it should be accessible, tailored and relevant to the stakeholder's life, while also being trustworthy. Additionally, scheduling and implementing renovations can be time-consuming, and some homeowners don't have enough time to deal with. It requires a suitable contractor to help the owners complete this series of complex steps. We therefore infer that energy retrofits is also influenced by perceived behavioral control, and postulate that:

**H3**: Energy retrofitting and intent to increase efficiency would be positively correlated with the degree of perceived control over the ability to retrofit.

- **Hypothesis 4 (Intent)**:

The principle proposition of the TPB is that behavior is the direct expression of intent. Intent (I) expresses a person's inclination or plan to actually perform a desired behavior. Many previous energy efficiency-related studies demonstrate this relationship/connection, so we infer that energy retrofit behavior is also influenced by intention [32] [33]. Therefore, we hypothesize:

**H4**: Energy retrofit behavior is positively associated with efficiency-enhancing intentions.

*2.1.2. Web tier*

The user interface (UI) is coded in HTML mark-up language and JavaScript. We use HTML to supplement the functions in UI with raw HTML to create highly customized applications. At this tier, users can perform various operations on the UI graphics system, including uploading and downloading datasets, performing selections, executing programs, displaying results and visualization [34].

As the main carrier of cross-platform and cross-device, the Web layer is used to handle HTTP requests, cross-communication, interactive operations, and some lightweight computing tasks. The program is developed on Google Chrome browser (Chrome DevTools), which is a highly integrated web-based application server with great compatibility for various platforms [35]. The PHP language is used to process requests. It can be selected to validate input data, ensuring connectivity between all tiers. The web tier can be seen as a tunnel that links the data flow between the front end and the back end, effectively presenting the results from the server. At the same time, the web tier is also an important display of this open digital ecosystem platform. Users can quickly and effectively obtain the results through the functions assembled by the website through user-centered interaction.

*2.1.3. Application tier*

The application layer is integrated with a series of programs, scripts and interactive data flows, running on the deployment platform managed by NodeJS-Express. NodeJS-Express is an open source platform that simplifies applications. It is a web application framework for Node.js, designed to build web applications and APIs [34]. Using this server, developers can host their applications in a controlled environment and they can also make apps available on the Internet. Algorithmic scripts written by developers are stored on Node.js. All scripts are developed based on JavaScript and Java programming language, and they are divided into different independent packages for different interactive functions. The specific algorithm of each function will be introduced in detail in the following subsection.

*2.1.4. Database tier*

The logic of the benchmark service is to generate reference standards based on the content of the database. That is to say, in addition to the built-in case data, users can also input their reference data independently to realize the interactive response based on the Web. Users perform calculations by inputting data to the front-end interface of the platform, and process the calculation data through the back-end server. For this reason, at the database tier, the stability and compatibility of the database has become an important consideration for developers. To this end, the MySQL open source relational database management system is applied [36]. For further scaling, using a document-oriented NoSQL database can be also considered.

On the other hand, use HTML5 to store user input data locally. Cookies can be used as a common storage method, but web storage needs to be more secure and fast, and these data will not be stored on the server. Based on this situation, we consider using Web Storage, the purpose of which is to overcome some limitations brought by cookies, that is, when the data needs to be strictly controlled on the client, there is no need to continuously send data back to the server.

*2.2. Application Components*

In application components, the primary purpose is to benchmark the user's building energy performance by comparing their input information with an appropriate reference. By classifying and dividing the data in the database, the following key influencing factors will be considered, including building location municipality, construction year of the building, number of families living in the building and total floor area of the building excluding the basement area and the corresponding Energy is used for consumption. The selection of influencing factors is based on the needs of the ongoing project, the availability of energy performance data and the knowledge of the assumed user input information.

All the corresponding influencing factors are constructed with user input parameters. Based on the user-centred design concept, the corresponding energy consumption including electricity usage (Input kWh) should be given in kWh or bill amount in Swedish Krona (SEK) and other energy sources (Input kWh). The output parameter in this system is the computation of the user's actual Energy Usage

Intensity (EUI). All the functions of the web-based system is illustrated in Fig 4. The summary of those function is shown in Tab 1.

In Function 1 (F1), the user input their electricity use in the number of bill amount, which is the user-centric method to simply the user's workload. However, due to the complexity of Swedish energy system [7], the building annual energy use (Input billtotal) is calculated using the following equation:

$$InputSEK = (SEK_{month} - SEK_{VAT} - SEK_{fee})/(SEK_{price} + SEK_{tax} + SEK_{network}) \qquad (1)$$

where the $SEK_{month}$ means the monthly bill from users; $SEK_{VAT}$ means Energy tax and VAT; $SEK_{fee}$ is a connection fuse charge depending on each fuse type and the data is taken from the main energy suppliers in Västerbotten region. $SEK_{price}$ is the pure electricity unit price excluding all charges of the user-specific group from the SCB database. $SEK_{tax}$ is energy tax per unit. $SEK_{network}$ is network charge per unit. There is one thing need to be mentioned, if the electricity supplier and grid operator are different, the terms of $SEK_{fee}$, $SEK_{tax}$, and $SEK_{network}$ can be neglected.

In function 4, the total energy usage $Input\ kWh_{total}$ includes user's energy consumption (F1/F2) and extra energy (F3). Energy Usage Intensity (EUI) can be calculated below:

$$EUI_{User} = \frac{Input\ kWh_{total}}{Input\ area_{m^2}} \qquad (2)$$

where $Input\ area_{m^2}$ is the total floor area excluding basement area.

In Function 6 (F6), Based on benchmark from database and user EUI value, the comparison can be displayed. All the numbers need to be compared with the range so the final scales can be generated, i.e, 5-points: *Very poor, Poor, Average, Good, Excellent*. 5-points results can provide advice to the users as reference for decision-making [37].

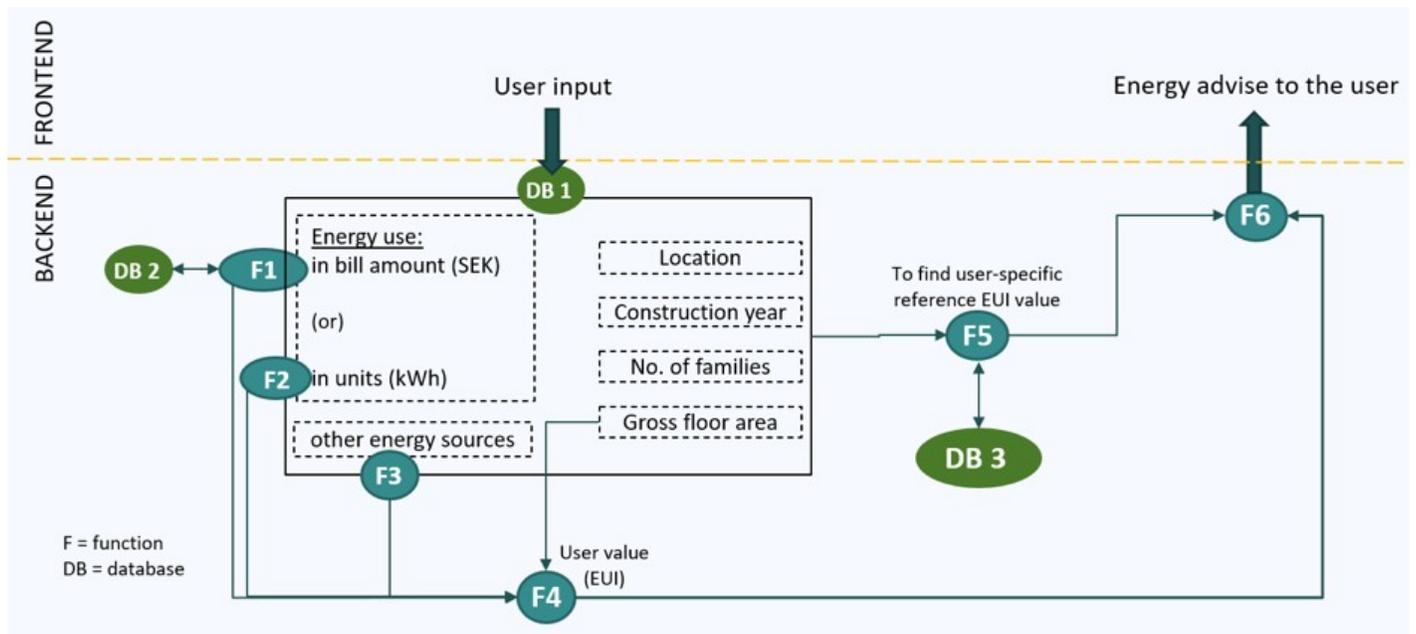

Figure 4: The functions of the web-based system

Table 1: The summary of those functions

| Functions | Database | Descriptions of Functions | Reference number |
|---|---|---|---|
| User input in bill amount (SEK) | DB2: Statistics Sweden | Receive the user's bill amount as input parameters and converts into kWh | F1 |
| User input in kWh | DB1: Users inputs | Receive the user's energy usage as input parameters and converts into kWh | F2 |
| Other energy sources and conversions | None | Other energy source e.g. Fuel oil, Natural gas, Firewood, Lignite Briquette etc. | F3 |
| User total value in EUI | None | Total energy usage including F1/F2 and F3. | F4 |
| Benchmark Reference | DB3: Boverket | Classification and analysis from dataset as references | F5 |
| Energy advice to the users | None | Comparing the benchmark with user's actual data | F6 |

## 2.3. Development workflow

Let us now focus on the user's domain. The data lineage and workflow diagram is illustrated in Fig 5.

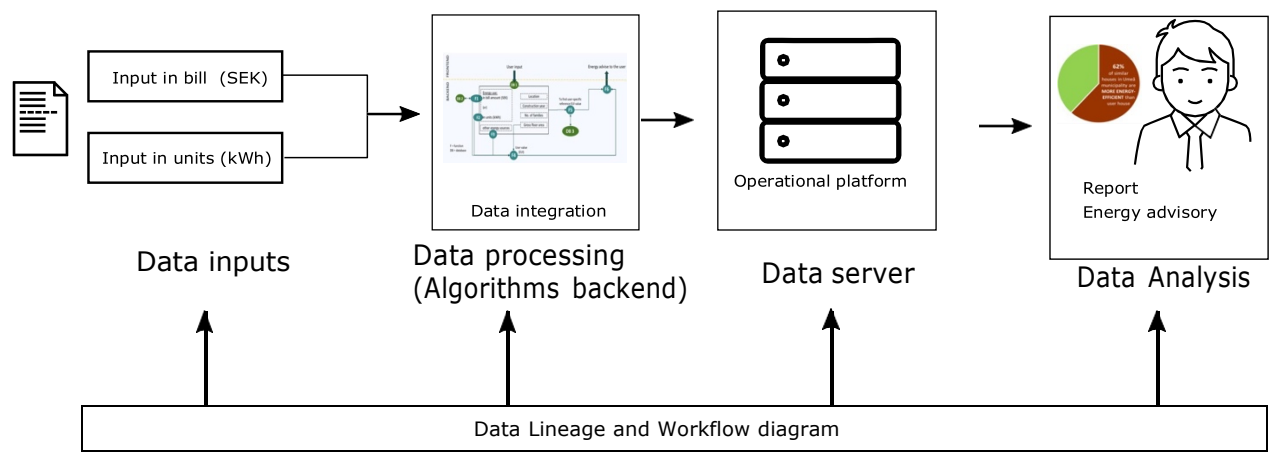

Figure 5: The data lineage and workflow diagram

The workflow of this web-based application with structural framework can be listed in the following steps:

- Step1: Input data
- Step2: Choosing options
- Step3: Computation on the server
- Step4: Get feedback from the output results

For users, step 1 is the most important part. Based on the user-centered concept is to allow users to achieve more effective results with more convenient and simple operations. We set different input parameters according to the actual situation of users, and they can obtain reference results more accurately through personalized selection via step 2.

With the complete data in hand, the next step should be to process it through a custom approach. Step 3 is the main function of our platform, and this is also the task that mainly needs the calculation and processing of the back-end server. We calculate users data through preset algorithms, and all calculation processes are traceable in our backend server. One thing also need to be mentioned, all data is anonymized in order to comply with the General Data Protection Regulation (GDPR) [38].

Step 4 occurs after the users submit their tasks, so they can immediately obtain a reference data as feedback. The obtained data mainly comes from the Boverket database and Statistics Sweden. The benchmarks given by these databases are very representative.

## 3. Quantitative Analysis

Quantitative analysis (QA) is a versatile approach that applies mathematical and statistical modeling, measurement, and research to understand mechanism behind behaviors [39]. The key output of quantitative analysts is representing a studied reality in terms of a numerical value. In this article, We use local and variance based sensitivity analysis (SA) to conduct our quantitative analysis and rank the importance of output contribution in terms of the model's inputs [40].

Variance-based sensitivity analysis (SA) is a statistical method used to identify the most important factors in a model or system that contribute to the output variability. This method is commonly used in quantitative analysis to understand the impact of different input parameters on the output of a model or system [41]. In variance-based SA, the input parameters are varied systematically to measure their impact on the output of the model or system. The analysis is based on the decomposition of the output variance into different components, each representing the contribution of a specific input parameter or combination of parameters. This method provides a rigorous and systematic approach to understanding the sensitivity of a model or system to different inputs. It can be particularly useful in cases where the input parameters are highly correlated or interact in complex ways. By using variance-based SA to conduct our quantitative analysis, we can gain insights into the relative importance of different input parameters on the output of our model or system. This information can be used to optimize the model or system by identifying the most important factors to focus on, or to inform policy decisions by identifying which factors have the greatest impact on the outcome of interest.

### 3.1. First-order sensitivity indices

Let's briefly discuss the Key Performance Indicators in the sensitivity analysis, which is the numerical value of the First-order sensitivity indices $S_i$ and total effect sensitivity indices $S_{T\,i}$. According to the definition of sensitivity analysis theory, the function $Y = f(X_1, X_2, \ldots, X_k)$ is shown as the response of a physical model. The first-order sensitivity indices are given as [42]

$$S_i = \frac{V_{x_i}\left[E_{X_{\sim i}}(Y \mid X_i)\right]}{V(Y)} \tag{3}$$

where The $V_{x_i}\left[E_{X_{\sim i}}(Y \mid X_i)\right]$ at the molecular position represents the main effect of $X_i$ on the output variable corresponding to the mathematical model. $V(Y)$ in the denominator position is taken as an unconditional variance of $Y$ and $V_{x_i}\left[E_{X_{\sim i}}(Y \mid X_i)\right]$ denotes the variance of the expected value of $Y$ conditional on $X_i$, i.e. $X_i$ is related and fixed to a value $X_{ji}$, $j = 1, \ldots, N$; with $N$ being a natural number whose value depends on the number of samples.

*3.2. Total effect sensitivity indices*

According to theoretical modeling, only the first-order sensitivity index is insufficient, since it is limited to the bias measurement of the model variance. Therefore, we need to extend the first-order to a higher-order coupling index reducing this part of the error. In this case, we define the total effect index as follows [42]

$$S_{T_i} = \frac{E_{x_{i_i}}[V_{X_i}(Y \mid X_{\sim i})]}{V(Y)} = 1 - \frac{V_{x_{\sim i}}[V_{X_i}(Y \mid X_{\sim i})]}{V(Y)} \tag{4}$$

where $V_{x_{\sim i}}[E_{X_i}(Y \mid X_{\sim i})]$ at the molecular position is the variance derived from of the expected value of all parameters $Y$ we need in this mode except for $X_i$, which subsequently denoted as $X_{\sim i}$. On the model output the first-order effect of $X_{\sim i}$ is represented by $V_{x_{\sim i}}[E_{X_i}(Y \mid X_{\sim i})]$, which the effect part of $X_i$ is not included in this item.

As we know, the overall effect of the input parameters $X_i$ on the output is expressed by the total effect index $S_{T_i}$, that is, the total effect is represented by the sum of the first-order term and all higher-order terms:

$$S_{T_i} = S_i + S_{i,\sim i} = 1 - S_{\sim i} \tag{5}$$

where $S_{\sim i}$ is a sum, including all parameters' sensitivity indices except of $i$.

The terms $S_i$ and $S_{T_i}$ in the equations [3] and [4] require a large number of samples to construct in order to achieve accuracy. The computational cost of applying these big data is usually high and large, especially when the amount of existing data is limited. Therefore, we construct a model $Y$ so-called a 'surrogate model' for approximating and representing the responses of this 'real' physical model.

*3.3. Surrogate Model*

In this subsection, the surrogate model from previous works [43][44][45][46][47] is an effective method. Here we briefly introduce two kinds of popular methods, namely polynomial regression method (with and without mixed terms) and moving least squares (MLS) approximation.

*3.3.1. Polynomial Regression model*

Given a general basic function can approximate the 'real' response $Y$ in the polynomial regression model (both linear and quadratic with mixed terms) is shown below:

$$\boldsymbol{P}_X^T(\boldsymbol{X}) = [X_1 \quad \ldots X_5 \ldots X_1^2 \quad \ldots X_5^2 \ldots X_1 X_2 \quad \ldots X_1 X_5 \ldots] \tag{6}$$

Subsequently, the SA method derives a sensitivity index for different input parameters. We integrate all vectors in to a vector $\boldsymbol{P}_X^T(\boldsymbol{X}) = [X_1 \quad \ldots X_k]$ according to the linear polynomial terms. $\hat{Y}$ is the response of the 'real' physical model $Y$ with k-parameters:

$$\widehat{Y} = \beta_0 + \sum_{i=1}^{K} \boldsymbol{\beta}_i P_X^T(X_i) + e \tag{7}$$

where the vector $\boldsymbol{\beta}$ can be obtained in minimizing the mean squared difference (MSD) S between the measured value at N training points in 'real' physical model. $x_j = [x_{j_1}, \ldots, x_{j_k}], j = 1, \ldots, N$ and the surrogate model's constructed value.

A least-square estimate (LSE) $\widehat{\boldsymbol{\beta}}$ is given below:

$$\widehat{\boldsymbol{\beta}} = (\boldsymbol{P}_X^T \boldsymbol{P}_X)^{-1} \boldsymbol{P}_X^T \boldsymbol{Y} \tag{8}$$

We also select coefficient of determination (COD) R2 and adjusted COD $R^2_{adj}$ as two metrics to estimate accuracy among all constructed models. All the metrics are shown by:

$$R^2 = 1 - \frac{\sum_{j=1}^{N} \left(Y_j - \hat{Y}_j\right)^2}{\sum_{j=1}^{N} \left(Y_j - \bar{Y}\right)^2} \tag{9}$$

$$R^2_{adj} = 1 - \frac{(N-1)}{(N-k_R)}(1 - R^2) \tag{10}$$

where N is the number of samples. The value of $k_R$ and $N$ are also considered into this measure method.

The main purpose of $R^2$ is to describe the similarity between those predicted and 'real' models. The adjusted COD $R^2_{adj}$ as an improvement includes more consideration on samples number.

Meanwhile, the quadratic without mixed terms basis function is also included in this research, mixed terms in Eq 6 can be removed:

$$P_X^T(X) = [X_1 \mid X_2 \quad ...X_k \quad X_1^2 \quad X_2^2 \quad ...X_k^2] \tag{11}$$

Since there is no mixed terms, Eq 7 can be modified into the corresponding form, given as follows:

$$\hat{Y} = \beta_0 + \sum_{i=1}^{K}\left(\beta_i X_i + \beta_{i_i} X_i^2\right) + e \tag{12}$$

*3.3.2. Moving Least Squares (MLS) regression model*

A local weighting functions is introduced to get a set of unordered point samples. This is an unique feature in Moving Least Squares (MLS) regression compare with other methods. The unknown coefficient can be computed by minimizing a difference value which is from the interpolation point $x$ in local approximation and the supporting point $x_i$ in input part of the physical model collection.

The basis of MLS function $\hat{Y}_{MLS}$ is presented as below:

$$\hat{Y}_{MLS} = P^T(X)\beta_w \tag{13}$$

where the quadratic terms are can shown in the polynomial basis **p**:

$$P_X^T(X) = [1 \quad X_1 \quad X_2 \quad ...X_k \quad X_1^2 \quad X_2^2 \quad ...X_k^2] \tag{14}$$

The weighted least squares coefficient $\beta_w$ can be obtained via minimizing a dicrete weighted $L^2$-norm $L_W$

$$L_W = \left(Y - p_{(X)}^T \beta_W\right)^T W(X)(Y - p^T(x)\beta_W) \tag{15}$$

with the unknown coefficients $\beta_w$ yielding

$$\frac{\partial L_W}{\partial \beta_w} = 0 \rightarrow \hat{\beta}_W = (X^T W(x)X)^{-1} X^T W(x) Y \tag{16}$$

Where $W(x)$ is a diagonal matrix

$$\begin{bmatrix} w(x-x_1) & 0 & ... & 0 \\ 0 & w(x-x_2) & ... & 0 \\ ... & ... & ... & ... \\ 0 & 0 & ... & w(x-x_N) \end{bmatrix} \tag{17}$$

in which the cubic polynomial weighting function that is applied in previous research [48]:

$$w(s) = \begin{cases} 1 - 3S^2 + 2S^3, & s \leq 1 \\ 0, & s > 1 \end{cases} \tag{18}$$

where hyperparameters here including D and s. D is the influential radius that is assumed to be a fixed value. $s = ||x - x_i||/D$ is a distance that is normalized between the supporting point N and the interpolation point;

Substituting Eq. 16 into Eq. 13 the MLS approximation can be presented as:

$$\hat{Y}_{MLS} = p^T(X)[X^TW(x)X]^{-1}X^TW(x)Y \tag{19}$$

*3.4. Variance-based method using Sobol'quasi-random sequences for computing Si and STi*

A Variance-based method using Sobol'quasi-random sequences is applied to compute the associated sensitivity index [49]. It is Monte Carlo estimation based on the Sobol's indices.

The construction of Sobol's indices is shown below. The matrix **A** and **B** are two independent samples with corresponding entries $a_{ji}$ and $b_{ji}$. the matrix is represented as ($N \times k$) and both of them are identical with same dimension for the input parameters **X**. Matrix $A_B^{(i)}$ $B_A^{(i)}$ are defined with entries from **A(B)** except the i-th column, which is taken from **B(A)** and compute $Si$ from **A**, $B_A^{(i)}$ or **B**, $A_B^{(i)}$ [50]:

$$V_{x_i}\left[E_{X_{\sim i}}(Y \mid X_i)\right] = \frac{1}{N}\sum_{j=1}^{N} f(A)_j f\left(B_A^{(i)}\right)_j - f_0^2 \tag{20}$$

where $(B)_j$ denotes the j-th row of matrix **B**. We rebuilt $S_{T_i}$ in Eqs 4 with $V_{x_{\sim i}}[E_{X_i}(Y \mid X_{\sim i})]$ as follows:

$$V_{x_{\sim i}}\left[E_{X_i}(Y \mid X_{\sim i})\right] = \frac{1}{N}\sum_{j=1}^{N} f(A)_j f\left(A_B^{(i)}\right)_j - f_0^2 \tag{21}$$

Similarly, the estimation of $S_i$ from Eqs 20 can be derived [51] [52]:

$$V_{x_i}\left[E_{X_{\sim i}}(Y \mid X_i)\right] = \frac{1}{N}\sum_{j=1}^{N} f(A)_j \left(f\left(B_A^{(i)}\right)_j - f(B)_j\right) \tag{22}$$

An appropriate improved method under the variance-based approach based on Quasi-Monte Carlo samples is proposed by Saltelli et al.[49], where the triplet $\mathbf{A}$, $\mathbf{B}$ and $A_B^{(i)}$ are used instead of $\mathbf{B}$, $\mathbf{A}$ and $\mathbf{B}_A^{(i)}$ in Eqs22. So the $S_{T_i}$ estimator in Eq 21 is obtained by [53]

$$V_{x_{\sim i}}[E_{X_i}(\mathbf{Y} \mid \mathbf{X}_{\sim i})] = \frac{1}{N}\sum_{j=1}^{N} f(\mathbf{A})_j \left( f(\mathbf{A})_j - f\left(A_B^{(i)}\right)_j \right) \tag{23}$$

Jansen et al. [54] proposed an alternative efficient approach for the estimators of can simplify the $S_i$ and $S_{T_i}$, where $V_{x_i}[E_{X_{\sim i}}(Y|X_i)]$ can be taken indirectly:

$$V_{x_i}\left[E_{X_{\sim i}}(\mathbf{Y} \mid X_i)\right] = \frac{1}{2N}\sum_{j=1}^{N} \left( f(\mathbf{B})_j - f\left(A_B^{(i)}\right)_j \right)^2 \tag{24}$$

After using $E_{x_{\sim i}}[V_{X_i}(Y|X_{\sim i})]$ in Eq 4 $S_{Ti}$ is computed by

$$V_{x_{\sim i}}\left[E_{X_i}(\mathbf{Y} \mid \mathbf{X}_{\sim i})\right] = \frac{1}{2N}\sum_{j=1}^{N} \left( f(\mathbf{A})_j - f\left(A_B^{(i)}\right)_j \right)^2 \tag{25}$$

Here we find applying a quasi-random number (QRN) too generate $\mathbf{A}$ and $A_B^{(i)}$ can simplify the computation, and then an estimate of $S_{T_i}$ can be obtained. In the article, the quasi-random sequences algorithm (QRSA) i.e. $LP\tau$ [55] is implemented in MATLAB through programming. It is applied to generate the samples $X_1, X_2, \ldots, X_k$ as uniformly as possible over the unit hypercube $\omega$. Three important requirements on $LP\tau$ sequences are listed [42]:

1. Increasing the sequence length and optimizing the uniformity of the distribution.
2. This distribution is also suitable for relatively small initial sets, i.e., the condition N is relatively small in value.
3. The computational complexity of this algorithm is acceptable.

After the quasi-random sequence with matrix size ($N \times 2k$) by QRSA method is ready, the whole structure is divided it into matrix **A** (left half) and **B** (right half) with matrix size ($N \times k$). The triplet **A**, **B** and $A_B^{(i)}$ include a higher amount of 'ideal' points. They are 'assigned' to the Triplet **A**, **B** and $B_A^{(i)}$ to compute $S_i$ and $S_{T_i}$.

### 3.5. Normalization of the input

Normalization is shrinking the data that needs to be processed to a certain range after processing (through a certain algorithm). Normalization can facilitate subsequent data processing, and can also ensure faster convergence when the program is running. The specific function of normalization is to summarize the statistical distribution of uniform samples. The purpose of normalization is to limit the preprocessed data to a specific range, so as to eliminate the adverse effects caused by singular sample data. It removes the unit constraints of the data and converts or changes them to dimensionless values, allowing better comparison and weighting of metrics in different units or different sizes [49].

In this case all the inputs should be normalized before constructing surrogate models. The i-th input factor $X_i$, the normalization produced by the s-th is shown as:

$$X_i^{(S),\,\text{norm}} = \frac{X_i^{(s)} - \min(X_i)}{\min X(X_i) - \max(X_i)} \qquad (26)$$

where $max(X_i)$ and $min(X_i)$ represent the maximum and minimum values of the i-th input sample, respectively.

## 4. Implementation

AURORAL is an EU Horizon 2020-funded project consisting of 25 organisations from 10 EU countries. The project focuses on delivering an interoperable, open, and digital middleware platform for rural ecosystem services in Europe, ultimately creating equal opportunities for rural and urban Europeans. The AURORAL middleware platform is demonstrated by cross-domain applications through 8 large- scale pilot regions in Europe including the Västerbotten region in Sweden [56]. The aim of the first step in Swedish pilot is to provide benchmarking of the users' building energy performance by comparing user input information with appropriate user reference peer groups.

This implementation is based on the AURORAL project. The schematic diagram of the benchmarking service as part of Swedish pilot is shown in Fig 6. Firstly, a reference database for benchmark standards was developed, which includes a large amount of data from Boverket (The National Board of Housing, Building and Planning), governing authority of Energy Performance Certificate (EPC) in Sweden [57]. A questionnaire was designed based on several basic hypotheses of the Theory of Planned Behavior, so related core issues are included as different questions. However, the purpose of an energy performance certificate (EPC) is to certify energy quality in a building given by an independent certified energy expert to the building owner when selling the property. EPC is valid for ten years, and consists of information about [58]:

- Location of the building
- Energy use for domestic hot water, electricity, heating and cooling in the building
- Energy use if it is a new building
- Total heating area
- Families numbers
- Ventilation and heating systems
- Energy performance rating from A to G, etc.

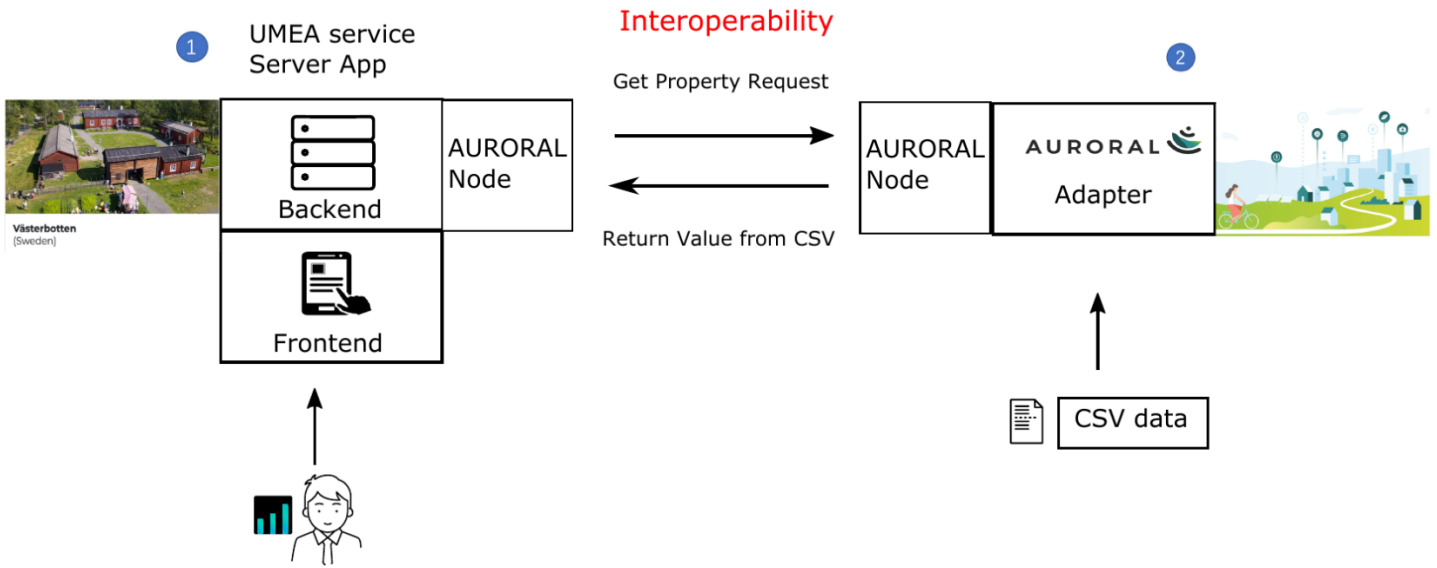

Figure 6: The schematic diagram of the Swedish pilot implementation

Boverket's EPCs dataset has 550,000 buildings, including residential, industrial, commercial, and public buildings. We focused more on houses built for one or two families since they are the most common type of housing in Sweden. This implementation determined the EPC of one/two-family buildings in the Västerbotten area. The 12,624 EPCs were classified according to four selected influencing factors, namely city, year of construction, number of resident households, and total floor area.

According to the previous research and the construction of the above dataset, we have screened out these influencing factors: S1: Construction year of the building; S2: Number of families living in the building; S3: Total floor area of the building excluding the basement area; S4: Energy usage; S5: Location of the building.

This implementation is based on the above-mentioned database and web-based application platform. It realized two functions, providing energy retrofits advisory and quantitative analysis.

Energy retrofits advisory has been realized through web-app interaction. By answering relevant questionnaires, users can give feedback based on the benchmark according to the situation of the building. Calculations to identify the user-specific situation based on their inputs and reference benchmark groups For instance, the advice will be visualized in the front-end that shall influence their decision-making. Thus, the advice given by the benchmark model is validated during the project testing period and further improved on feedback received. The advice is given by answering the questions:

- *Does this user building need to be renovated or not to meet Sweden's national energy efficiency target (SEET) 2050?*

By submitting a questionnaire, users can get feedback as shown in the Fig 7

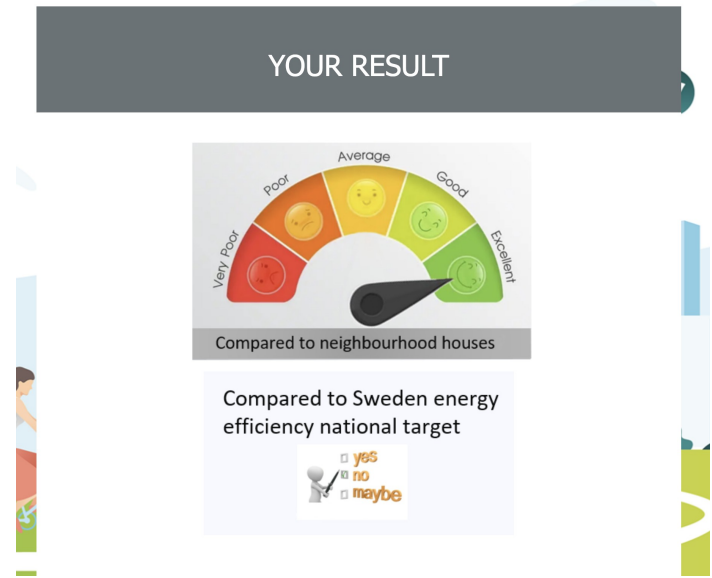

Figure 7: The results from energy advisory platform

Advice: No (since the user actual EUI < allowed EUI in 2022 and the user EUI is classified in Excellent scale)

- *How is the user building performance compared to similar neighbourhood buildings?*

Advice: The feedback is illustrated in Fig 8

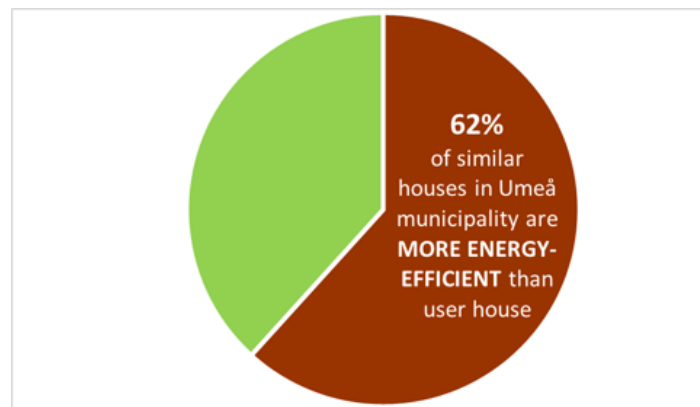

Figure 8: The results from energy advisory platform compared to similar neighbourhood buildings

The second important part in this implementation is the quantitative analysis, through the sensitivity computation of different influencing factors to obtain the contribution of input factors to the output results.

Since the limited amount of questionnaire data, both the questionnaire data and Boverket data for analysis were integrated. A total of 3182 groups of raw data is ready for further data processing. We construct surrogate models on the original data, and then generate 100,000 sets for sensitivity analysis, including first-order indices and total-effect indices, respectively. The results are shown in the Tab 2.

From the results we know that S4 and S3 are more important to the overall results. Then it is followed by S2, and finally by S1 and S5. From a numerical point of view, the importance of S1 and S5 is very small and can be ignored. The Figs 9 and 10 show these trends in first-order indices and total-effect indices respectively.

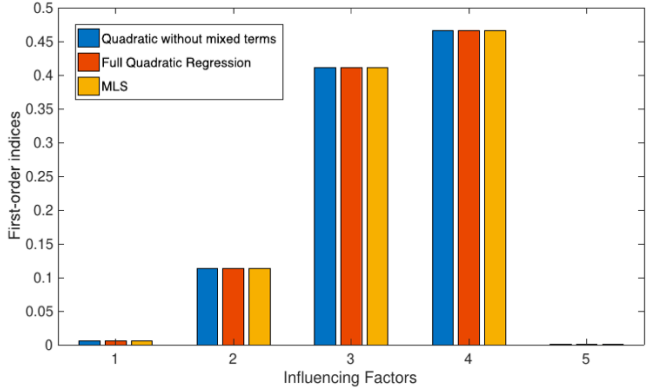
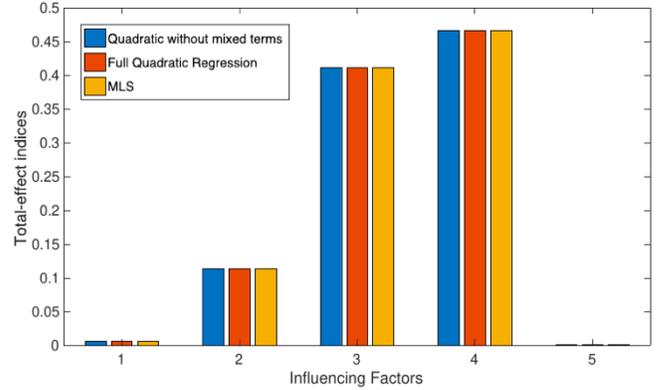

Figure 9: First-order indices   Figure 10: Total-effect indices

Further analysis of the sensitivity indices obtained by different surrogate models is almost consistent, indicating that the ranking of the factors is referenced to a certain extent. The first-order sensitivity index of a single variable is also consistent with the global sensitivity index, indicating the selected variables in this physical model are all independent variables without obvious interaction among them.

These quantitative data can also be well explained from the results, because in the definition of EUI, energy consumption and housing area are important components. The number of household members directly affects energy use and therefore indirectly affects the EUI. Although the construction year and latitude position are also the performance of the data, they have limited influence on the results because they cannot be reflected in the constitutive equation.

Based on the results we obtained and the information provided, there are a few policy implications that can be drawn. Firstly, energy efficiency policies should consider household size: the fact that household size directly affects energy use and, consequently, the EUI, suggests that energy efficiency policies should take into account the number of household members in their design. For example, incentives could be provided for households with smaller sizes to encourage energy-efficient practices. Secondly, energy consumption should be targeted in housing policies: since energy consumption is a critical component of the EUI, policies that aim to reduce energy consumption in housing, such as improving insulation and promoting energy-efficient appliances, can be effective in lowering EUI. Lastly, While the construction year and latitude position can affect energy use to some extent, their impact is limited compared to other factors such as energy consumption and household size. For example, even if a building is constructed with energy-efficient materials and techniques, its occupants' behavior can still have a significant impact on energy consumption. Besides, the construction year and latitude position is determined by various factors such as local building codes, economic conditions, climate and topography. These factors are difficult to regulate through policy measures.

## 5. Conclusions

The focus of this study is to introduce a web-based decision-making support system platform specifically designed for homeowners to benchmark their buildings. With rising energy prices in Europe and increased carbon emissions globally, there is a pressing need to improve energy efficiency in the building sector. The proposed open ecosystem framework aims to provide advice to stakeholders in the Västerbotten region by quantifying influencing factors and addressing key questions. The platform utilizes a 4-tier system architecture programmed for user-centric interactive designs and visualization accessible via a web browser. To accomplish this, several big data-driven methods have been integrated into the backend server as functions. The data collection process began with a questionnaire based on the Theory of Planned Behavior (TPB). Reference data for the benchmark is obtained from Boverket and SCB and is integrated into the framework as APIs. To ensure user-centered design, input can be given in either the number of units consumed (kWh) or the electricity bill amount (in SEK) via a connected API. This benchmark model has been implemented in the AURORAL project in the Västerbotten region. The platform's energy renovation advisory and quantitative analysis capabilities have been utilized to verify the ecosystem framework platform's service.

Table 2: First-order and total effects sensitivity indices computed on different surrogate models

| **Influencing factors** | **Quadratic without mixed terms** | **Full Quadratic Regression** | **MLS** |
|---|---|---|---|
| First-order indices $\hat{S}_i$ | First-order indices $\hat{S}_i$ | First-order indices $\hat{S}_i$ | First-order indices $\hat{S}_i$ |
| $\hat{S}_1$: Construction year | $\hat{S}_1$=0.0064 | $\hat{S}_1$=0.0064 | $\hat{S}_1$=0.0064 |
| $\hat{S}_2$: Family Members | $\hat{S}_2$=0.1141 | $\hat{S}_2$=0.1141 | $\hat{S}_2$=0.1141 |
| $\hat{S}_3$: Total area of house | $\hat{S}_3$=0.4111 | $\hat{S}_3$=0.4111 | $\hat{S}_3$=0.4112 |
| $\hat{S}_4$: Energy usage | $\hat{S}_4$=0.4671 | $\hat{S}_4$=0.4671 | $\hat{S}_4$=0.4671 |
| $\hat{S}_5$: Location (latitude and longitude) | $\hat{S}_5$=0.0012 | $\hat{S}_5$=0.0012 | $\hat{S}_5$=0.0012 |
|  | $\sum_{i=1}^{5} \hat{S}_i$=0.9999 | $\sum_{i=1}^{5} \hat{S}_i$=0.9999 | $\sum_{i=1}^{5} \hat{S}_i$=1 |
| Total-effect indices $\hat{S}_{Ti}$ | Total-effect $\hat{S}_{Ti}$ | Total-effect indices $\hat{S}_{Ti}$ | Total-effect indices $\hat{S}_{Ti}$ |
| $\hat{S}_{T1}$: Construction year | $\hat{S}_{T1}$=0.0064 | $\hat{S}_{T1}$=0.0064 | $\hat{S}_{T1}$=0.0064 |
| $\hat{S}_{T2}$: Family Members | $\hat{S}_{T2}$=0.1141 | $\hat{S}_{T2}$=0.1141 | $\hat{S}_{T2}$=0.1141 |
| $\hat{S}_{T3}$: Total area of house | $\hat{S}_{T3}$=0.4111 | $\hat{S}_{T3}$=0.4111 | $\hat{S}_{T3}$=0.4112 |
| $\hat{S}_{T4}$: Energy usage | $\hat{S}_{T4}$=0.4671 | $\hat{S}_{T4}$=0.4671 | $\hat{S}_{T4}$=0.4671 |
| $\hat{S}_{T5}$: Location (latitude and longitude) | $\hat{S}_{T5}$=0.0012 | $\hat{S}_{T5}$=0.0012 | $\hat{S}_{T5}$=0.0012 |
|  | $\sum_{i=1}^{5} \hat{S}_{Ti}$=0.9999 | $\sum_{i=1}^{5} \hat{S}_{Ti}$=0.9999 | $\sum_{i=1}^{5} \hat{S}_{Ti}$=1 |

The framework is characterized by the following features: 1) Implementation of a web-based and user-friendly application; 2) flexible integration of APIs, databases, algorithms, and server functions; 3) Technical barrier-free access for users to get helpful insights into choosing the right EEMs and finding a credible renovation company from the benchmarking module of the energy advisory service.

This open digital ecosystem framework is based on the AURORAL project middleware platform and is currently being used effectively in the Västerbotten region, the northern part of Sweden. This AURORAL ecosystem is for the digital transformation of smart communities and rural areas [59], including the Sustainability of smart rural mobility and tourism, etc [60]. The improved energy advisor service, enabled by the developed platform, can significantly reduce the cost of decision-making, enabling decision-makers to participate in such professional knowledge-required decisions in a deliberate and efficient manner [61]. Users can obtain fast and accurate energy advisory and solve professional problems. At the company level, this platform can be used to implement precise services and provide overall improvement advice and implementation. For society, this digital platform can increase energy efficiency, reduce carbon emissions, and mitigate the impact of climate change to a certain extent.

The study's results suggest several policy implications. Energy efficiency policies should consider household size, as it directly affects energy use and the EUI. Incentives could be provided for households with smaller sizes to encourage energy-efficient practices. Energy consumption should be targeted in housing policies, with a focus on reducing energy consumption through measures such as improving insulation and promoting energy-efficient appliances. However, the impact of construction year and latitude position on energy use is limited compared to energy consumption and household size, and policy measures to regulate them are difficult due to various factors such as building codes, economic conditions, climate, and topography.

This study focuses on user-centric energy retrofits in Northern Sweden, where energy efficiency and retrofitting are crucial concerns for rural communities, as they often have older buildings and infrastructure that can benefit from modernization to reduce energy consumption, lower costs, and decrease environmental impact. The study's findings could shed light on the effectiveness of data-driven approaches to energy retrofits, which may be particularly relevant for rural regions or city communities with unique energy challenges. The integration of an open digital ecosystems platform indicates a move towards creating a smart community. Smart communities leverage technology and data to enhance the quality of life for residents, improve resource management, and promote sustainable development. By studying the case in Northern Sweden, the research might identify specific strategies and tools that can be applied to other rural areas aspiring to become smart communities. This study's emphasis on data-driven quantitative analysis highlights the importance of using data and evidence-based approaches in addressing issues. This can be particularly valuable for both rural and city communities, as they often face energy efficiency issues and need to make informed decisions to allocate resources effectively. By demonstrating the value of data-driven decision-making in the context of energy retrofits, the study may inspire similar practices in other rural development initiatives. The concept of an integrated open digital ecosystems platform suggests the use of interconnected digital solutions and services. Such ecosystems can foster collaboration, innovation, and knowledge-sharing among different stakeholders in rural areas, including local governments, businesses, and residents. Understanding the benefits and challenges of implementing such platforms in a rural setting could be highly relevant for promoting holistic rural development. Meanwhile, the study's case in Northern Sweden could serve as a model or inspiration for similar projects in other rural regions. Rural areas worldwide share common challenges and opportunities, and successful approaches from one region can often be adapted and applied elsewhere. The research may provide insights into factors that contribute to the success of energy retrofit projects in rural settings, making it easier to replicate or scale up in other communities.

In summary, the study's relevance to rural issues and smart communities lies in its focus on energy retrofits, data-driven analysis, digital ecosystems, and potential applicability to similar rural contexts. The findings could contribute to more sustainable and efficient rural development and offer valuable lessons for policymakers, planners, and stakeholders working on rural revitalization and smart community initiatives.

**Acknowledgment**

We gratefully acknowledge the support of the EU project H2020-AURORAL Grant agreement ID: 101016854 (Architecture for Unified Regional and Open digital ecosystems for Smart Communities and Rural Areas Large scale application) and the Kempe Foundation Sweden (Kempestiftelserna - Stiftelserna J.C. Kempes och Seth M. Kempes minne). This work is also funded and supported by J. Gust. Richert stiftelse, SWECO, Sweden (Grant agreement ID: 2023-00884).

The computations handling were enabled by resources provided by the Swedish National Infrastructure for Computing (SNIC) and Academic Infrastructure for Supercomputing in Sweden (NAISS) at High-Performance Computing Center North (HPC2N) partially funded by the Swedish Research Council through grant agreement no. 2018-05973 and no. 2022-06725.